# *AISCliteracy*: Assessing Artificial Intelligence and Cybersecurity Literacy Levels and Learning Needs of Students

Devendra Chapagain, Auckland Univ. of Technology, Auckland, New Zealand, devendra.chapagain@bimc.tu.edu.np
Naresh Kshetri, Rochester Institute of Technology, Rochester, New York, USA, naresh.kshetri@rit.edu
Bishwo Prakash Pokharel, Sault College, Toronto, Ontario, Canada, bishwo889@hotmail.com

**Abstract**

Artificial intelligence (AI) is rapidly transforming global industries and societies, making AI literacy an indispensable skill for future generations. While AI integration in education is still emerging in Nepal, this study focuses on assessing the current AI literacy levels and identifying learning needs among students in Chitwan District of Nepal. By measuring students' understanding of AI and pinpointing areas for improvement, this research aims to provide actionable recommendations for educational stakeholders. Given the pivotal role of young learners in navigating a rapidly evolving technological landscape, fostering AI literacy is paramount. This study seeks to understand the current state of AI literacy in Chitwan District by analyzing students' knowledge, skills, and attitudes towards AI. The results will contribute to developing robust AI education programs for Nepalese schools. This paper offers a contemporary perspective on AI's role in Nepalese secondary education, emphasizing the latest AI tools and technologies. Moreover, the study illuminates the potential revolutionary impact of technological innovations on educational leadership and student outcomes. A survey was conducted to conceptualize the newly emerging concept of AI and cybersecurity among students of Chitwan district from different schools and colleges to find the literacy rate. The participants in the survey were students between grade 9 to 12. We conclude with discussions of the affordances and barriers to bringing AI and cybersecurity education to students from lower classes.

## 1. Introduction:

Artificial intelligence (AI) is rapidly transforming our world, impacting various aspects of our lives from daily routines to complex decision-making processes. AI technologies are increasingly being integrated into diverse fields, including healthcare, finance, education, and transportation. As AI continues to evolve, it becomes critical for individuals to possess a fundamental understanding of its capabilities, limitations, and potential societal implications. This understanding, often referred to as AI literacy, empowers individuals to interact effectively with AI technologies, make informed decisions, and contribute meaningfully to the development and application of AI in an ethical and responsible manner.

AI has rapidly permeated various sectors, and education is no exception. Traditionally perceived as a domain of supercomputers, AI has evolved to encompass embedded computer systems, with its algorithms and systems increasingly integrated into educational processes [1]. As the educational landscape transforms, researchers are exploring sophisticated AI techniques such as deep learning and data mining to address complex pedagogical challenges and personalize instruction for individual learners.

In the past, technology in education was mainly used to give students information. Teachers were like instructors, passing knowledge to students. However, things changed. People started to understand that learning is more than just getting information. It's about thinking and creating new ideas. So, technology began to be used in a different way. Instead of just giving information, it started to help students build their own understanding. This meant using technology to encourage students to think critically, solve problems, and work together. Basically, the goal shifted from teachers telling students what to think to helping students learn to think for themselves.

The integration of AI in education is increasingly seen as a means to cultivate 21st century skills. [2] Researchers emphasize the need for AI systems that prioritize student-centered learning experiences, promoting active knowledge construction and adaptability to individual needs. This shift from technology-centric to learner-focused approaches is crucial for realizing the full potential of AI in education.

Artificial Intelligence in Education (AIEd) emerges as a complex interplay of computer science, statistics, education, and increasingly, cybersecurity. This interdisciplinary field seeks to harness computational power to enhance and optimize learning processes while ensuring the security and privacy of educational data. By integrating principles from cognitive psychology, neuroscience, and cybersecurity, AIEd aims to develop intelligent systems that can adapt to individual learner needs, facilitate knowledge acquisition, and safeguard against emerging cyber threats [3]. The convergence of these disciplines has spawned subfields such as Educational Data Mining, Learning Analytics, and Cybersecurity Education, which collectively contribute to a more robust and secure educational landscape. Preparing the next generation of the workforce requires equipping youth with the knowledge and skills necessary to thrive in the AI-driven era, including a strong foundation in cybersecurity [4].

## 2. Problem statement

The rapid advancement and integration of artificial intelligence (AI) across various sectors globally underscores the necessity for a workforce that is not only technically skilled but also AI-literate. In this context, AI literacy refers to the understanding of AI concepts, applications, ethical implications, and cybersecurity concerns that are essential for responsible and informed interaction with AI technologies. However, in developing countries like Nepal, the educational system has yet to fully incorporate AI and cybersecurity education, leaving a significant gap in students' preparedness to navigate a world increasingly dominated by these technologies.

Nepal's current educational curriculum does not adequately address AI and cybersecurity topics, especially at the secondary education level. This gap is particularly concerning in regions like Chitwan District, which is growing very fast, but students are at a disadvantage in terms of exposure to and understanding of emerging technologies. Despite the growing presence of digital technologies in Nepalese society, there is a lack of empirical data on the current state of AI literacy among students. Without this foundational knowledge, students may find themselves ill-equipped to engage with or benefit from AI innovations, potentially exacerbating the digital divide.

Research by Iqbal, Khan, and Imran [5] revealed that AI literacy is essential for students to effectively engage with and benefit from AI technologies. Their study emphasized that without adequate AI education, students are less likely to develop the critical thinking skills necessary to navigate the ethical and societal implications of AI. Similarly, Tuomi [6] argued that AI literacy encompasses not only technical understanding but also the ability to critically assess the social impacts of AI technologies, which is crucial for empowering students as active participants in a digital society.

Moreover, the lack of AI and cybersecurity education poses a broader societal risk. As AI becomes more pervasive, individuals without basic literacy in these areas may become passive consumers of technology rather than active, informed participants in its development and application. This situation could lead to various negative outcomes, including vulnerability to cyber threats, misinformation, and an inability to critically assess the implications of AI-driven decisions in their personal and professional lives.

The specific challenges in Chitwan District highlight the urgent need to assess the current levels of AI and cybersecurity literacy among students and to identify the learning needs that must be addressed to close these gaps. The absence of such literacy not only hinders individual students' prospects but also impedes the region's ability to keep pace with technological advancements that are transforming global economies and societies.

This study, therefore, seeks *to address these gaps by conducting a detailed assessment of AI and cybersecurity literacy levels among students in Chitwan District*. It aims *to identify specific areas where students lack understanding and to provide actionable insights that can inform the development of targeted educational strategies*. By doing so, this research will contribute to the broader goal of preparing Nepalese students for the challenges and opportunities of a rapidly evolving technological landscape.

## 3. Literature Review

AI literacy encompasses understanding AI concepts, applications, and ethical considerations. Previous studies highlight the importance of integrating AI education at an early age to prepare students for future challenges. However, there is limited research on AI literacy in the context of developing countries like Nepal. This review explores existing literature on AI education, focusing on methodologies, outcomes, and identified gaps. We also examine Nepal's educational policies and initiatives related to AI, revealing a need for comprehensive strategies to improve AI literacy.

Knox, J. in [7] his paper highlighted that China's AI development is characterized by a dynamic interplay between state-led strategic planning and market-driven innovation. The educational sector is at the nexus of these forces, serving as both a catalyst for AI research and a target for AI applications.

Ouyang and Jiao [8] categorize the integration of AI in education into three primary paradigms. The first, AI-directed, positions students as passive recipients of AI-delivered instruction. The second paradigm, AI-supported, involves a more collaborative approach where AI acts as a tool to facilitate student learning. Finally, the AI-empowered paradigm envisions students as active agents in their education, with AI serving as an enabler for creative and independent learning. This progression reflects the evolving role of AI in education, moving from a teacher-centered to a learner-centered approach.

Research conducted by Chen et al. [9] revealed an increasing interest in using AI for educational purposes within the academic community. The field of AI in education (AIEd) has demonstrated potential across various applications. These include providing student support, enabling data-driven decision-making, creating dynamic classroom environments, enhancing student engagement, and facilitating efficient instruction through personalized and adaptive learning experiences.

AI literacy, as described by Yi [10], encompasses more than just the technical understanding of AI systems. It involves a comprehensive grasp of the social implications of technology, basic knowledge of AI, and the ability to effectively use technology in various contexts. This literacy is essential for individuals to maintain their independence rights in a society increasingly shaped by AI. Without this understanding, there is a risk that individuals may become passive subjects in an AI-driven world, rather than active participants who can shape how AI impacts their lives and society.

Moreover, the benefits of AI technology are not automatically realized; they depend on the capacity of individuals to seek out, understand, and utilize the information provided by AI systems effectively. Therefore, appropriate education is necessary to equip individuals with the skills to obtain the desired information from AI technologies and to navigate the complex ethical, social, and political landscapes they influence.

AI has emerged as a transformative force in education, with the potential to revolutionize teaching and learning practices. By offering personalized instruction through adaptive learning systems, AI enables educators to cater to the unique needs of each student [11]. This individualized approach fosters deeper engagement and improved learning outcomes. Moreover, AI's capabilities extend beyond personalized learning. Research has demonstrated its utility in assessment, evaluation, and predictive analytics, empowering educators to make data-driven decisions and optimize educational interventions [12].

As primary educators of online safety, teachers face the challenge of navigating a complex digital environment where appropriate behaviors are not always evident. To effectively protect teens from cyber threats, it is imperative to enhance their cybersecurity, privacy, and digital literacy skills [13]. Concurrently, teachers require comprehensive cybersecurity training and resources to equip them for this critical role.

## 4. Importance of AI and Cybersecurity education in the 21st century

AI literacy encompasses a range of skills and knowledge, including the ability to engage in adaptive learning, critically analyze information, and develop a deep understanding of technology. It empowers individuals to harness the transformative potential of AI, fostering innovation and problem-solving. Conversely, a lack of AI literacy can create a digital divide, where certain individuals or groups are excluded from the benefits of AI technology. This exclusion can lead to challenges such as information inequality, difficulty in using AI-powered devices, and a limited ability to fully participate in the digital age. The demand for AI skills is rapidly increasing. Proficiency in AI can open doors to careers in data analytics, cybersecurity, and other fields, with industry leaders like Google, Microsoft, and IBM actively seeking AI talent. Additionally, a strong foundation in cybersecurity complements AI expertise, enhancing career prospects.

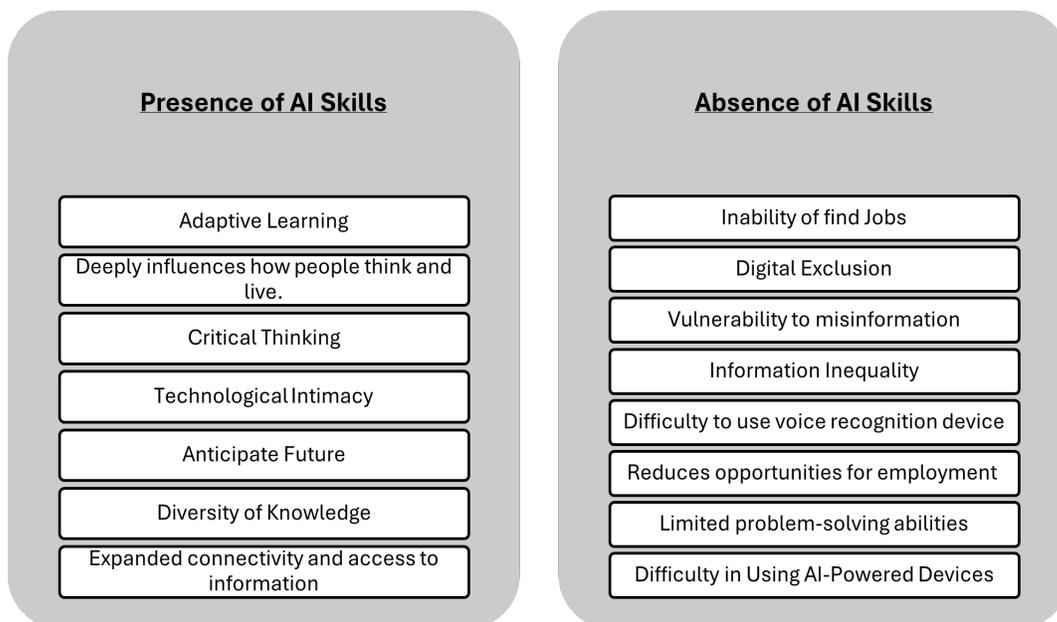

Figure 1: Impact of AI literacy

Similarly, cybersecurity education is essential in today's interconnected world to protect individuals and societies from the growing threats posed by online activities. It is crucial to introduce cybersecurity education early, as children are increasingly exposed to the internet from an early age [14]. While the internet offers vast opportunities, it also presents significant risks, such as cyberbullying, online scams, and identity theft.

Beyond defending against external threats, cybersecurity education plays a key role in addressing issues like internet addiction. Excessive online gaming and gambling can have detrimental effects on individuals and families. By understanding cybersecurity principles, people can develop the skills needed to manage their online behavior and prevent harmful dangers that arise in the digital realm.

Education is undergoing a change in thinking from mere knowledge transmission to cultivating skilled, innovative individuals. Modern education emphasizes developing students' ability to leverage technology for problem-solving and societal contributions. AI is a pivotal tool in this transformation, enabling higher education institutions to adapt to the digital age [15]. By analyzing vast datasets on job markets, curricula, and student profiles, AI can optimize educational pathways, aligning student skills with industry demands. This data-driven approach fosters a dynamic ecosystem where institutions can effectively connect students with suitable career opportunities.

The opinion scale on artificial intelligence in education revealed a strong positive sentiment among teachers. A substantial majority expressed confidence in AI's capacity to enhance student learning through personalized instruction and efficient content delivery [16]. Teachers overwhelmingly recognized AI's potential to streamline administrative tasks, thereby affording them more time for student interaction and professional development.

## 5. Educational policies, AI, and Cybersecurity initiatives in Nepal

As Nepal navigates the challenges and opportunities presented by the digital age, integrating artificial intelligence (AI) and cybersecurity into its educational framework has become increasingly important. However, the country's progress in these areas has been gradual, reflecting broader socio-economic and infrastructural limitations that characterize many developing nations. Nepal's education policies have undergone significant changes in recent years, with a growing emphasis on integrating information and communication technology (ICT) into the educational system. The Education Act of 2016 restructured the education system, transitioning from a three-tier to a two-tier model and introducing the Secondary Education Examination (SEE) for 12th graders.[17] While progress has been made, challenges remain in fully integrating ICT into Nepal's education system. Issues such as limited internet connectivity, lack of teacher training, and inadequate infrastructure need to be addressed to ensure effective ICT integration.

### 5.1 Current Education Policies

Nepal's educational policies have historically focused on expanding access to basic education, improving literacy rates, and addressing gender disparities. However, the incorporation of advanced technological subjects like AI and cybersecurity into the national curriculum is still in its initial stages. The Ministry of Education, Science, and Technology (MoEST) has recognized the need to modernize the curriculum to keep pace with global trends, but implementation has been slow due to resource constraints and a lack of technical expertise.

In 2019, the Government of Nepal launched the "National Education Policy 2076", which emphasizes the integration of information and communication technology (ICT) in education. This policy laid the groundwork for introducing AI and cybersecurity concepts at various educational levels. However, the policy lacks specific guidelines or frameworks for integrating these subjects into the curriculum, leaving schools to develop their own approaches [18].

Furthermore, the "Nepal ICT in Education Master Plan 2013-2017", was an earlier effort to introduce ICT education across Nepalese schools. This plan included distributing computers to schools, training teachers in basic ICT skills, and developing digital content for students. While AI and cybersecurity were not the primary focus, this initiative laid the foundation for future efforts in these areas [19]. While ICT-related policies have been developed since 2009, implementation challenges persist. Infrastructure limitations, lack of skills and training [20], and ineffective government practices have hindered the widespread adoption of ICT in education.

**5.2 AI and Cybersecurity Initiatives**

Despite the slow progress in policy development, there have been several notable initiatives aimed at enhancing AI and cybersecurity education in Nepal. These initiatives are driven by private sector collaborations, non-governmental organizations (NGOs), and international partnerships.

For instance, the organization *Karkhana* has been instrumental in introducing AI concepts to school children through workshops and hands-on learning experiences. Similarly, *Nepal Research and Education Network (NREN)* has been actively involved in promoting cybersecurity awareness through seminars, online courses, and collaborations with international experts.

A significant government-led initiative is the *Cybersecurity Awareness Assessment* conducted by the Nepal Telecommunications Authority (NTA) in collaboration with international partners. This report highlights the growing need for cybersecurity education in Nepal and outlines strategies for raising awareness among students and educators [21]. AI significantly enhances cybersecurity by improving threat detection, response times, and predictive insights. A prime example is its application in intrusion detection systems [22]. AI algorithms can efficiently analyze vast datasets, identifying patterns indicative of malicious activity and adapting to emerging threats in real time. This initiative-taking approach is crucial for combating the ever-evolving landscape of cybercrime.

## 6. Methodology

This study employs a mixed-methods approach, combining quantitative surveys and qualitative interviews to assess AI literacy levels and learning needs among students in Chitwan District. We selected a sample of three hundred and three students from schools in Chitwan District using stratified random sampling. Data collection involved a structured questionnaire assessing AI knowledge and skills to students. Ethical considerations included informed consent and data confidentiality. The Likert scale responses were analyzed to identify trends and patterns in AI literacy levels and learning needs.

The data analysis involved calculating the Relative Importance Index (RII) for each element to determine their significance and frequency. A Likert scale was used to assess respondents' opinions, with scores ranging from 1 (Strongly Disagree) to 5 (Strongly Agree). The RII was calculated as the weighted average of responses, with higher RII values indicating greater importance.

## 7. Findings and Discussion

Of the 303 students surveyed in Chitwan District, Nepal, 55.12% were male and 44.88% were female. Internet access was widespread, with 97.69% of students reporting easy access from home. However, a small percentage (1.98%) faced difficulties in accessing the internet.

Smartphones were the primary device used for online activities by 66.34% of students. Laptops/computers were used by 26.70%, and tablets by 1.65%. The remaining 5.27% utilized other devices. The following table presents an analysis of students' understanding, awareness, and interest in Artificial Intelligence (AI) and Cybersecurity. The results offer insights into the current state of AI and cybersecurity literacy among the student population. In the survey carried out in Chitwan District, Nepal, 303 students were studied out of which 55.12% were male and 44.88% were female.

Table 1: RII Scores for AI and Cybersecurity Literacy

| SN | AI and Cybersecurity Knowledge Questions | RII | RANKS |
|---|---|---|---|
|  | **Artificial Intelligence (AI)** |  |  |
| 1 | I have a good understanding of what AI is | 0.78020 | 3 |
| 2 | I can give examples of how AI is used in everyday life. | 0.78548 | 4 |
| 3 | I am familiar with different types of AI technologies (machine learning, robotics). | 0.72739 | 6 |
| 4 | I am aware of the potential benefits and risks of AI. | 0.82970 | 2 |
| 5 | I believe AI will have a positive impact on my future. | 0.72277 | 1 |
| 6 | I am interested in learning more about AI. | 0.83696 | 5 |
| 7 | I can identify AI applications in various industries (healthcare, finance, education). | 0.77360 | 6 |
| 8 | I have used/interacted with AI-powered devices/apps (virtual assistants, recommendation systems). | 0.76370 | 7 |
| 9 | I am comfortable discussing AI topics with my peers or teachers. | 0.80528 | 8 |
| 10 | My school provides resources or courses on AI education. | 0.65215 | 9 |
|  | **Cybersecurity** |  |  |
| 11 | I am aware of common cybersecurity threats such as phishing, malware, and hacking. | 0.87855 | 10 |
| 12 | I regularly use strong, unique passwords for my online accounts. | 0.84554 | 11 |
| 13 | My school provides adequate education on how to stay safe online. | 0.75050 | 12 |
| 14 | I can recognize suspicious emails or messages that may be phishing attempts. | 0.82574 | 13 |
| 15 | I use antivirus software to protect my devices. | 0.74323 | 14 |
| 16 | I have been a victim of a cyber-attack (knowingly or unknowingly). | 0.50825 | 15 |
| 17 | I have completed/attended a cybersecurity training or workshop. | 0.45149 | 16 |
| 18 | I do often change my passwords for online accounts. | 0.74785 | 17 |
| 19 | I regularly update the software or operating systems on devices. | 0.77426 | 18 |

### 7.1. AI Literacy:

The study's results reveal that students have a moderate to high understanding of AI, with key indicators showing a strong foundational knowledge. As reflected in Table 1, the statement *"I have a good understanding of what AI is"* had an RII of 0.78020, ranking third in AI-related questions. This suggests that a majority of students believe they understand AI at a basic level. Similarly, *"I can give examples of how AI is used in everyday life"* had an RII of 0.78548, highlighting that students can recognize AI applications in real-world contexts.

However, when asked about familiarity with advanced AI concepts, such as machine learning and robotics, the RII dropped slightly to 0.72739, indicating some knowledge gaps in more technical aspects. This is consistent with their comfort discussing AI topics with peers or teachers, where the RII was 0.80528, suggesting confidence in casual conversation but not necessarily in-depth technical discussions.

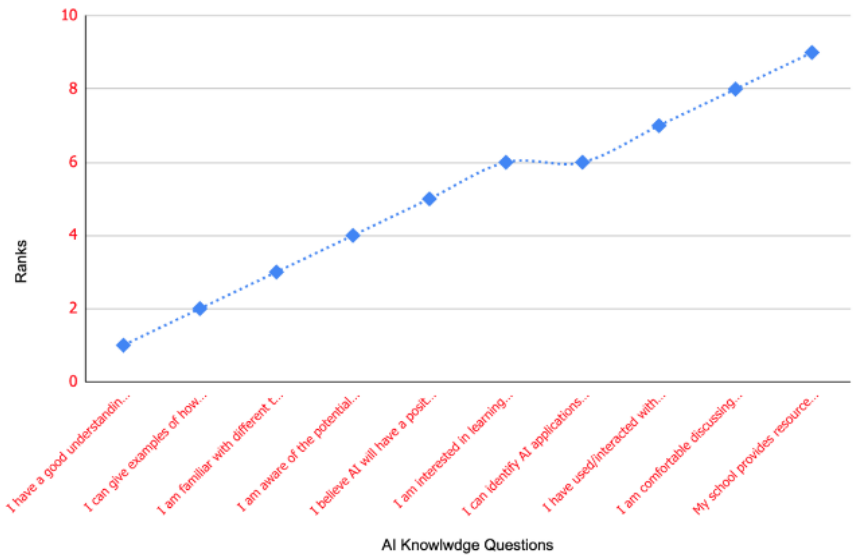

Figure 2: Ranks vs. AI knowledge questions plot

In terms of awareness, students demonstrated a clear understanding of the benefits and risks associated with AI, as indicated by the relatively high RII of 0.82970 for *"I am aware of the potential benefits and risks of AI"*. This awareness is further reflected in the positive outlook students have on AI, with an RII of 0.72277 for *"I believe AI will have a positive impact on my future"*, indicating optimism about AI's role in their lives and future careers. Moreover, a notable proportion of students are eager to learn more about AI, with an RII of 0.83696 for *"I am interested in learning more about AI"*, ranking this statement as one of the highest overall, reflecting a keen interest in expanding their AI literacy.

**7.2. Cybersecurity Literacy:**

The findings on cybersecurity awareness and literacy suggest that students have a reasonably strong grasp of cybersecurity concepts. The highest-ranked statement, *"I am aware of common cybersecurity threats such as phishing, malware, and hacking"*, had an RII of 0.87855, indicating a high level of awareness regarding cybersecurity threats. Similarly, students displayed good cybersecurity practices, as shown by the RII of 0.84554 for *"I regularly use strong, unique passwords for my online accounts"*.

Despite these positive findings, the data reveals a concerning lack of formal cybersecurity education in schools. The statement *"My school provides adequate education on how to stay safe online"* had an RII of 0.75050, ranking lower than expected, indicating that although students are aware of key cybersecurity threats, they may not be receiving sufficient formal education on how to manage these risks.

Additionally, the results show a lack of cybersecurity training experiences, with *"I have completed/attended a cybersecurity training workshop"* receiving one of the lowest RII values (0.45149), suggesting that few students have participated in formal cybersecurity programs. This highlights a potential area for intervention in the education system to provide more structured and comprehensive cybersecurity education.

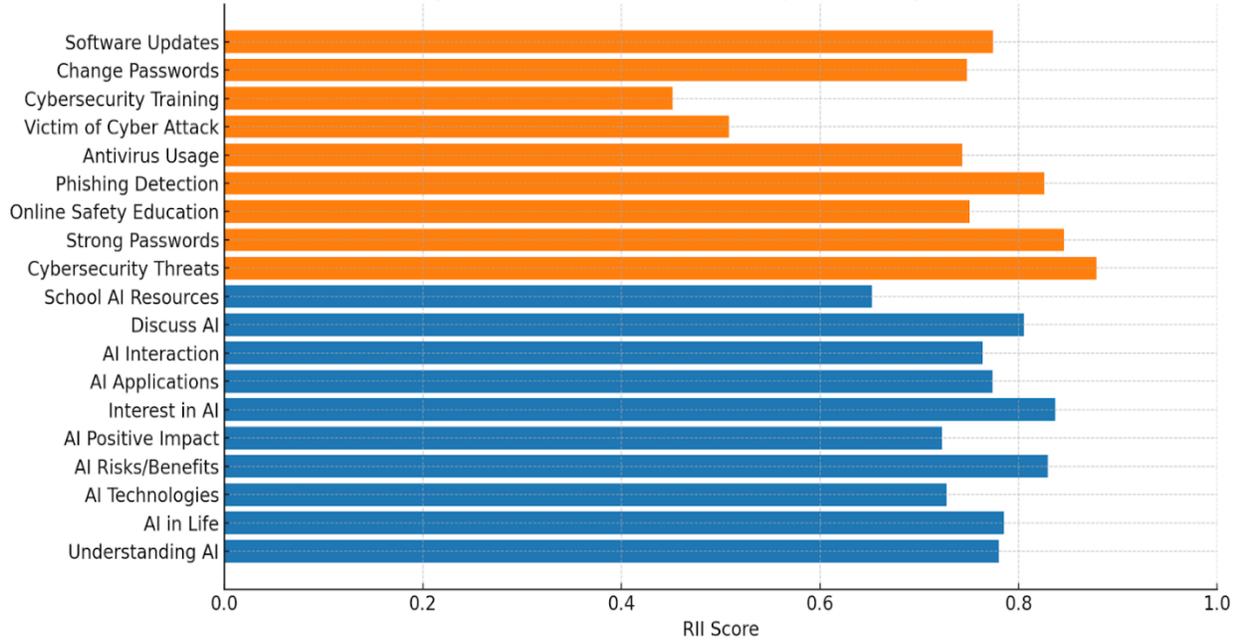

Figure 3: RII Scores for AI and Cybersecurity Literacy

**7.3. Analysis of AI and Cybersecurity Literacy in Context**

The data also reveal a strong relationship between students' access to AI resources in school and their overall understanding of AI. Schools that provide resources or courses on AI education were associated with higher levels of understanding, as reflected in the RII of 0.65215 for *"My school provides resources or courses on AI education"*. However, the relatively low ranking of this statement suggests that many schools are not offering sufficient AI educational resources, which could be limiting students' ability to fully develop their AI literacy.

Furthermore, there was no significant difference in AI literacy between students who had internet access at home and those who did not, indicating that while internet access is important for self-directed learning, it may not be a key factor in determining overall AI literacy levels. This suggests that formal education, rather than self-learning via the internet, plays a more critical role in shaping students' understanding of AI concepts.

**7.4. Discussion**

The findings of this study highlight both opportunities and challenges in AI and cybersecurity education in the Chitwan District, Nepal. AI Literacy levels are moderate to high, with students showing a strong foundational understanding of AI concepts and a positive outlook towards AI's role in their future. This optimism and eagerness to learn more about AI provide a valuable opportunity for educational institutions to expand their curriculum to include more comprehensive AI education, particularly in advanced topics like machine learning and robotics, where knowledge gaps still exist.

On the other hand, the results raise concerns about the availability of formal educational resources for both AI and cybersecurity. The relatively low RII for statements like *"My school provides resources of courses on AI education"* and *"My school provides adequate education on how to stay safe online"* suggests that schools are not adequately equipped for fostering a deep understanding of AI or cybersecurity. This finding aligns with previous studies that emphasize the need for integrating AI and cybersecurity education into school curricula [19][21]. Expanding school programs to include AI and Cybersecurity courses could significantly enhance students' literacy in these areas.

Cybersecurity Literacy, while high in terms of awareness, reveals critical gaps in practical education and training. Despite students' understanding of cybersecurity threats, the lack of formal training programs or workshops limits their ability to apply their knowledge effectively. The low RII for *"I have completed/attended a cybersecurity training or workshop"* suggests that experiential learning in cybersecurity is minimal. As such, policymakers and educators should consider implementing more hands-on cybersecurity training, particularly given the increasing importance of cybersecurity in today's digital landscape [18].

## 8. Future Scope

The study, while offering valuable insights into AI literacy in Chitwan District, is limited by its scope and reliance on self-reported data. Future research should broaden its scope and incorporate qualitative methods to gain a deeper understanding. Analyzing the curriculum and evaluating educational programs will inform the development of targeted interventions. Furthermore, exploring ethical implications is crucial for responsible AI use. By addressing these limitations, future research can contribute to a more comprehensive understanding of AI literacy in Nepal and inform effective educational strategies to prepare students for the digital age.

**References**
[1] L. Chen, P. Chen, and Z. Lin, "Artificial Intelligence in Education: A Review," in *IEEE Access*, vol. 8, pp. 75264-75278, 2020, doi: 10.1109/ACCESS.2020.2988510
[2] Lameras P, Arnab S. Power to the Teachers: An Exploratory Review on Artificial Intelligence in Education. *Information*. 2022; 13(1):14. https://doi.org/10.3390/info13010014
[3] Xieling Chen, Haoran Xie, Di Zou, Gwo-Jen Hwang, Application and theory gaps during the rise of Artificial Intelligence in Education, Computers and Education: Artificial Intelligence, Volume 1, 2020, 100002, ISSN 2666-920X, https://doi.org/10.1016/j.caeai.2020.100002.
[4] Lee, I., Ali, S., Zhang, H., DiPaola, D., & Breazeal, C. (2021, March). Developing middle school students' AI literacy. In *Proceedings of the 52nd ACM technical symposium on computer science education* (pp. 191-197).
[5] Iqbal, M., Khan, N. U., & Imran, M. (2024). The role of Artificial Intelligence (AI) in transforming educational practices: Opportunities, challenges, and implications. *Qlantic Journal of Social Sciences*, 5(2), 348–359. https://doi.org/10.55737/qjss.349319430
[6] Tuomi, I. (2023). Beyond mastery: Toward a broader understanding of AI in Education. *International Journal of Artificial Intelligence in Education*, 34(1), 20–30. https://doi.org/10.1007/s40593-023-00343-4
[7] Knox, J. (2020). Artificial intelligence and education in China. Learning, Media and Technology, 45(3), 298–311. https://doi.org/10.1080/17439884.2020.1754236
[8] Ouyang, F., & Jiao, P. (2021). Artificial intelligence in education: The three paradigms. Computers and Education: Artificial Intelligence, 2, 100020. doi:10.1016/j.caeai.2021.100020
[9] Chen, X., Zou, D., Xie, H., Cheng, G., & Liu, C. (2022). Two Decades of Artificial Intelligence in Education: Contributors, Collaborations, Research Topics, Challenges, and Future Directions. Educational Technology & Society, 25(1), 28–47. https://www.jstor.org/stable/48647028
[10] Yi, Y. (2021). Establishing the concept of AI literacy: Focusing on competence and purpose. JAHR-–European Journal of Bioethics, 12 (2), 353–368.
[11] Huang, J., Saleh, S., & Liu, Y. (2021). A review on artificial intelligence in education. *Academic Journal of Interdisciplinary Studies*, *10*(3).
[12] Derinalp, P. Past, Present, and Future of Artificial Intelligence in Education: A Bibliometric Study. Sakarya University Journal of Education, 14(2 (Special Issue-Artificial Intelligence Tools and Education)), 159-178.
[13] Maqsood, S., & Chiasson, S. (2021). Design, development, and evaluation of a cybersecurity, privacy, and digital literacy game for tweens. ACM Transactions on Privacy and Security (TOPS), 24(4), 1-37.
[14] Rahman, N. A. A., Sairi, I. H., Zizi, N. A. M., & Khalid, F. (2020). The importance of cybersecurity education in school. International Journal of Information and Education Technology, 10(5), 378-382.
[15] Alshahrani, B. T., Pileggi, S. F., & Karimi, F. (2024). A Social Perspective on AI in the Higher Education System: A Semi systematic Literature Review. *Electronics*, *13*(8), 1572.
[16] Uygun, D. (2024). Teachers' perspectives on artificial intelligence in education. Advances in Mobile Learning Educational Research, 4(1), 931-939. https://doi.org/10.25082/AMLER.2024.01.005

[ 17] Karki, H. (2019). A brief history of public education, information & communication technology (ICT) and ICT in public education in Nepal. *Deerwalk Journal of Computer Science and Information Technology*, 78-103.

[18] Government of Nepal. (2019). National Education Policy 2076. https://kms.pri.gov.np/dams/pages/view.php?ref=3882&search=%21collection1805&k=0e27cc7aaa

[19] Nepal ICT in education master plan, 2013-2017 | ICT in education ... (n.d.-b). https://www.ictedupolicy.org/resource-library/content/nepal-ict-education-master-plan-2013-2017

[20] Joshi, D. R. (2017). Policies, practices and barriers of ICT utilization in school education in Nepal. *International Journal of Research in Social Sciences*, *7*(2), 408-417.

[21] Cybersecurity Awareness Assessment Report for Nepal. (n.d.-a). https://nta.gov.np/uploads/contents/Cybersecurity-Awareness-Report-2015.pdf

[22] Zeadally, S., Adi, E., Baig, Z., & Khan, I. A. (2020). Harnessing artificial intelligence capabilities to improve cybersecurity. *Ieee Access*, *8*, 23817-23837.